\documentclass[12pt]{article}
\usepackage{graphicx}
\usepackage{amssymb}
\usepackage{cite}

\begin{document}

\def\qq{\langle \bar q q \rangle}
\def\uu{\langle \bar u u \rangle}
\def\dd{\langle \bar d d \rangle}
\def\sp{\langle \bar s s \rangle}
\def\GG{\langle g_s^2 GG \rangle}
\def\Tr{\mbox{Tr}}
\def\figt#1#2#3{
        \begin{figure}
        $\left. \right.$
        \vspace*{-2cm}
        \begin{center}
        \includegraphics[width=10cm]{#1}
        \end{center}
        \vspace*{-0.2cm}
        \caption{#3}
        \label{#2}
        \end{figure}
	}
	
\def\figb#1#2#3{
        \begin{figure}
        $\left. \right.$
        \vspace*{-1cm}
        \begin{center}
        \includegraphics[width=10cm]{#1}
        \end{center}
        \vspace*{-0.2cm}
        \caption{#3}
        \label{#2}
        \end{figure}
                }

\def\ds{\displaystyle}
\def\beq{\begin{equation}}
\def\eeq{\end{equation}}
\def\bea{\begin{eqnarray}}
\def\eea{\end{eqnarray}}
\def\beeq{\begin{eqnarray}}
\def\eeeq{\end{eqnarray}}
\def\ve{\vert}
\def\vel{\left|}
\def\ver{\right|}
\def\nnb{\nonumber}
\def\ga{\left(}
\def\dr{\right)}
\def\aga{\left\{}
\def\adr{\right\}}
\def\lla{\left<}
\def\rra{\right>}
\def\rar{\rightarrow}
\def\nnb{\nonumber}
\def\la{\langle}
\def\ra{\rangle}
\def\ba{\begin{array}}
\def\ea{\end{array}}
\def\tr{\mbox{Tr}}
\def\ssp{{\Sigma^{*+}}}
\def\sso{{\Sigma^{*0}}}
\def\ssm{{\Sigma^{*-}}}
\def\xis0{{\Xi^{*0}}}
\def\xism{{\Xi^{*-}}}
\def\qs{\la \bar s s \ra}
\def\qu{\la \bar u u \ra}
\def\qd{\la \bar d d \ra}
\def\qq{\la \bar q q \ra}
\def\gGgG{\la g^2 G^2 \ra}
\def\q{\gamma_5 \not\!q}
\def\x{\gamma_5 \not\!x}
\def\g5{\gamma_5}
\def\sb{S_Q^{cf}}
\def\sd{S_d^{be}}
\def\su{S_u^{ad}}
\def\sbp{{S}_Q^{'cf}}
\def\sdp{{S}_d^{'be}}
\def\sup{{S}_u^{'ad}}
\def\ssp{{S}_s^{'??}}

\def\sig{\sigma_{\mu \nu} \gamma_5 p^\mu q^\nu}
\def\fo{f_0(\frac{s_0}{M^2})}
\def\ffi{f_1(\frac{s_0}{M^2})}
\def\fii{f_2(\frac{s_0}{M^2})}
\def\O{{\cal O}}
\def\sl{{\Sigma^0 \Lambda}}
\def\es{\!\!\! &=& \!\!\!}
\def\ap{\!\!\! &\approx& \!\!\!}
\def\ar{&+& \!\!\!}
\def\ek{&-& \!\!\!}
\def\kek{\!\!\!&-& \!\!\!}
\def\cp{&\times& \!\!\!}
\def\se{\!\!\! &\simeq& \!\!\!}
\def\eqv{&\equiv& \!\!\!}
\def\kpm{&\pm& \!\!\!}
\def\kmp{&\mp& \!\!\!}
\def\mcdot{\!\cdot\!}
\def\erar{&\rightarrow&}

% .........................................................

\def\simlt{\stackrel{<}{{}_\sim}}
\def\simgt{\stackrel{>}{{}_\sim}}

% .........................................................

\title{
         {\Large
                 {\bf
Magnetic and quadrupole moments of light spin--1 mesons in
light cone QCD sum rules
                 }
         }
      }

\author{\vspace{1cm}\\
{\small T. M. Aliev \thanks
{e-mail: taliev@metu.edu.tr}~\footnote{permanent address:Institute
of Physics,Baku,Azerbaijan}\,\,,
A. \"{O}zpineci \thanks
{e-mail: ozpineci@p409a.physics.metu.edu.tr}\,\,,
M. Savc{\i} \thanks
{e-mail: savci@metu.edu.tr}} \\
{\small Physics Department, Middle East Technical University,
06531 Ankara, Turkey} }

\date{}

\begin{titlepage}
\maketitle
\thispagestyle{empty}

\begin{abstract}
The magnetic and quadrupole moments of the light--vector and axial--vector
mesons are calculated in the light cone QCD sum rules. Our results for 
the static properties of these mesons are compared with the predictions of 
lattice QCD as well as other approaches existing in the literature.  
\end{abstract}

%\vspace{1cm}
~~~PACS number(s): 11.55.Hx, 13.40.Em, 14.20.Jn
\end{titlepage}

\section{Introduction}

The electromagnetic form factors of the mesons and baryons represent an
important tool for understanding their internal structure in terms of quarks
and gluons. Investigation of the electromagnetic form factors of the
nucleons, both theoretically and experimentally, is one of the main research
areas in particle physics \cite{R9501,R9502,R9503}. The electromagnetic 
form factors of the pseudoscalar mesons, especially the pion, has been 
extensively studied (see \cite{R9504,R9505,R9506,R9507,R9508,R9509} 
and references therein).
Unfortunately, the form factors of the vector mesons have received less 
interest (see \cite{R9510,R9511,R9512,R9513,R9514,R9515} for recent studies). 
The electromagnetic form factors of vector mesons is also the subject of 
recent lattice QCD calculations (see \cite{R9509,R9510} and \cite{R9516}). 
In the present work we calculate the magnetic and quadrupole moments of 
the light--vector and axial--vector mesons in light cone QCD sum rule (LCSR) 
(for more about light cone QCD sum rule method, see \cite{R9517} and 
\cite{R9518}). Note that the magnetic moment of $\rho$ meson is calculated 
in this framework in \cite{R9515}. Here in this work we present improved 
calculations for the $\rho$ meson including the distribution amplitudes 
which are neglected in \cite{R9515}, as well as a new result on quadrupole
moment of the $\rho$ meson Furthermore, it should be noted that, the 
magnetic moment of $\rho$ mesons calculated in the framework of QCD sum 
rules using the external field technique in \cite{R9511}.

The paper is organized as follows. In section 2 the LCSR for the magnetic
and quadrupole moments of the light--vector and axial--vector mesons
are derived. Section 3 is devoted to the numerical analysis. Furthermore, this
section contains out conclusions and comparison of our results with the ones
obtained from lattice QCD calculations.
    
\section{Light cone QCD sum rules for the magnetic and quadrupole moments of
light--vector and axial--vector mesons}

In this section we derive the LCSR for the magnetic and quadrupole moments
of the light--vector and axial--vector mesons. For this aim we consider the
correlation function of the two vector meson currents in the presence of the
external electromagnetic field, which is the main object in LCSR.
\bea
\label{e9501}
\Pi_{\mu\nu} = i \int d^4x e^{ipx}\lla 0 \vel {\cal T} \{J_\nu(x) 
J_\mu^\dagger (0)\} \ver 0 \rra _\gamma~, 
\eea
where $\gamma$ is the external electromagnetic field, $J_\nu
(x)=\bar{q}_1(x) \Gamma_\nu q_2(x)$ is the interpolating current of the
light--vector (axial--vector) mesons when
$\Gamma_\nu=\gamma_\nu~(\gamma_\nu\gamma_5)$.

According to QCD sum rules method, the correlation function is calculated in
two different ways:
\begin{itemize}
\item in terms of quark degrees of freedom interacting with the
nonperturbative QCD vacuum (theoretical part),
\item being saturated by the mesons (as is in our case) having the same
quantum as the interpolating current (phenomenological part).
\end{itemize}

We start our analysis by calculating the phenomenological part of the
correlation function. Inserting a complete set states with the same quantum
numbers as the interpolating current and isolating the ground state meson,
we get:

\bea
\label{e9502}
\Pi_{\mu\nu} = {\lla 0 \vel J_\nu \ver i(p) \rra \lla i(p) \ve  i(p^\prime)
\rra_\gamma \lla i(p^\prime) \vel J_\mu^\dagger (0) \ver 0 \rra 
\over (p^2-m_i^2) (p^{\prime 2}-m_i^2)} + \cdots~,
\eea
where $i$ represents light--vector or axial--vector mesons, $p^\prime=p+q$, $q$
is the photon momentum and the dots correspond to the contribution
of higher states and continuum.

It follows from Eq. (\ref{e9502}) that in order to calculate the
phenomenological part of the correlation function, the matrix elements $\lla
0 \vel J_\nu \ver i(p) \rra$ and $\lla i(p) \ve  i(p^\prime)
\rra_\gamma$ are needed. The matrix element $\lla 0 \vel J_\nu \ver i(p)
\rra$ is defined as:
\bea
\label{e9503}
\lla 0 \vel J_\nu \ver i(p) \rra = f_i m_i~.
\eea

Imposing the parity and time reversal invariance of the electromagnetic 
interaction, the electromagnetic vertex of the light--vector (axial--vector)
is described in terms of the three form factors in the following way
\cite{R9519}: 
\bea
\label{e9504}
\lla i(p,\varepsilon^r) \ve  i(p^\prime,\varepsilon^{r\prime}) \rra_\gamma
\es - \varepsilon^\rho (\varepsilon^r)^\alpha (\varepsilon^{r\prime})^\beta
\Big\{ G_1 (Q^2) g_{\alpha\beta} (p+p^\prime)_\rho + G_2 (Q^2) (q_\alpha
g_{\rho\beta} - q_\beta g_{\rho\alpha}) \nnb \\
\ek {1\over 2 m_i^2} G_3 (Q^2)
q_\alpha q_\beta (p+p^\prime)_\rho \Big\}~,
\eea
where $\varepsilon^\rho$ id the photon and $(\varepsilon^r)^\alpha$,
$(\varepsilon^{r\prime})^\beta$ are the light--vector (axial--vector) meson
polarization vectors. The covariant form factors $G_1$, $G_2$ and $G_3$ can
be expressed in terms of the Sachs charge, magnetic and quadrupole form
factors as follows \cite{R9519,R9520}:
\bea
\label{e9505}
F_{\cal C} \es G_1(Q^2) + {3\over 3}\eta F_{\cal D} (Q^2)~, \nnb \\
F_{\cal M} \es G_2(Q^2) ~, \nnb \\
F_{\cal D} \es G_1(Q^2) - G_2(Q^2) + (1+\eta) G_3(Q^2)~,
\eea
where $\eta=Q^2/4 m_i^2$.

The charge $q_i$, magnetic moment $\mu_i$ and quadrupole moment ${\cal D}_i$ 
are determined from $F_{\cal C}$, $F_{\cal M}$ and $F_{\cal D}$, respectively,
at zero momentum transfer,  
\bea
\label{e9506}
e F_{\cal C}^i (0) \es q_i~, \nnb \\
e F_{\cal M}^i (0) \es 2 m_i \mu_i ~, \nnb \\
e F_{\cal D}^i (0) \es m_i^2 {\cal D}_i~.
\eea

Substituting the expressions in Eqs. (\ref{e9503}) and (\ref{e9504}) into
Eq. (\ref{e9502}), and performing summation over polarizations of the 
light--vector (axial--vector) meson, for the phenomenological part of the
correlation function we have:
\bea
\label{e9507}
\Pi_{\mu\nu}^{ph} \es f_i^2 m_i^2 {\varepsilon^\rho \over (m_i^2 - p^2)
[m_i^2 - (p+q)^2]} \Bigg\{ G_1(Q^2) (p+p^\prime)_\rho \Bigg[ g_{\mu\nu} -
{p_\mu p_\nu \over m_i^2} - {p_\mu^\prime p_\nu^\prime \over m_i^2} \nnb \\
\ar {p_\mu^\prime p_\nu \over 2 m_i^4} (Q^2 + 2 m_i^2) \Bigg]
+ G_2(Q^2) \Bigg[ q_\mu g_{\nu\rho} - q_\nu g_{\mu\rho} - {p_\nu \over
m_i^2} \Bigg( q_\mu p_\rho - {1\over 2} Q^2 g_{\mu\rho} \Bigg) \nnb \\
\ar {p_\mu^\prime \over m_i^2} \Bigg( q_\nu p_\rho^\prime + {1\over 2} 
Q^2 g_{\nu\rho} \Bigg)
-  {p_\mu^\prime p_\nu p_\rho \over m_i^4} Q^2 \Bigg]
- {1 \over m_i^2} G_3(Q^2) (p+p^\prime)_\rho \Bigg[ q_\mu q_\nu -
{p_\mu q_\nu \over 2 m_i^2} Q^2 \nnb \\
\ar {p_\mu^\prime q_\nu \over 2 m_i^2} Q^2 -
{p_\mu^\prime p_\nu \over 4 m_i^4} Q^4 \Bigg] \Bigg\}~,
\eea
where $Q^2 = -q^2$.   

As has already been noted, in order to determine the magnetic and quadrupole
moments, the values of the form factors are needed only at $Q^2=0$.
Substituting Eq. (\ref{e9505}), as well as the relations $p^\prime = p+q$ and
$q\varepsilon = 0$ into Eq. (\ref{e9507}), we obtain the final result for
$\Pi_{\mu\nu}$:
\bea                               
\label{e9508}
\Pi_{\mu\nu}^{ph} \es f_i^2 m_i^2 {\varepsilon^\rho \over (m_i^2 - p^2)
[m_i^2 - (p+q)^2]} \Bigg\{ 2 F_{\cal C} (0) p_\rho \Bigg[ g_{\mu\nu} -
{p_\mu p_\nu \over m_i^2} - {p_\mu q_\nu \over m_i^2} \Bigg] \nnb \\
\ar F_{\cal M} (0) \Bigg[ q_\mu g_{\nu\rho} - q_\nu g_{\mu\rho} - {p_\rho \over
m_i^2} (p_\mu q_\nu - p_\nu q_\mu) \Bigg] - [F_{\cal C} (0) + F_{\cal D} (0)]
{p_\rho \over m_i^2} q_\nu q_\mu \Bigg\}~.
\eea

In determining magnetic and quadrupole magnetic moments, different
structures are needed. For this purpose, we prefer to choose the structures
which do not get contribution from the contact terms after Borel
transformation (more about contact terms can be found in \cite{R9521}). The
structures $q_\mu \varepsilon_\nu$ and $q_\nu q_\mu (\varepsilon p)$ do not
receive contributions from the contact terms, and for this reason we choose
them for extracting the magnetic and quadrupole magnetic moments of the
light--vector (axial--vector) mesons.

Our next task is the calculation of the correlation function from the QCD
side in terms of the photon distribution amplitudes. Using the explicit
expression of the interpolating current in $x$ representation, for the
correlation function we get
\bea                               
\label{e9509}
\Pi_{\mu\nu}^{th} = i \int d^4x e^{ipx} \lla \gamma(q) \vel S(x) \Gamma_\mu S(-x)
\Gamma_\nu \ver 0 \rra ~.
\eea

The correlation function contains three different combinations:
\begin{itemize}
\item a) perturbative contributions,
\item b) ``mixed contribution", i.e., photon interacts with the quark propagator
perturbatively, and other quark fields organize the quark condensate,
\item c) nonperturbative contribution, i.e., photon is emitted at long
distance.
\end{itemize}

In calculating the contribution coming from (c), the propagator of the quark
field is expanded near the light cone $x^2=0$, as a result of which appears
the matrix elements of the nonlocal operators $\lla \gamma(q) \vel \bar{q}
(x_1) \Gamma q(x_2) \ver 0 \rra$, between the vacuum and one photon states,
i.e., these matrix elements are expressed in terms of the photon
distribution amplitudes (Da's).     
 
In Eq. (\ref{e9509}), $S(x)$ is the full propagator of the light quark
expanded in the light cone up to linear order in the quark mass
\cite{R9522,R9523},
which has the following form:
\bea
\label{e9510}
S(x) \es {i \rlap/x \over 2 \pi^2 x^4} - {m_q \over 4 \pi^2 x^2} - {\qq \over
12} \Bigg( 1- {i m_q\over 4} \rlap/x \Bigg) - {x^2 \over 192 \pi^2} m_0^2
\qq \Bigg( 1- {i m_q\over 6} \rlap/x \Bigg) \nnb \\
\ek i g_s \int_0^1 du \Bigg\{ {\rlap/x
\over 16 \pi^2 x^2} G_{\mu\nu} (ux) \sigma^{\mu\nu} - 
{i\over 4 \pi^2 x^2} u x^\mu G_{\mu\nu} (ux) \gamma^\nu (ux) \nnb \\
\ek {i m_q \over 32 \pi^2} G_{\mu\nu}
(ux) \sigma^{\mu\nu} \Bigg[ \ln \Bigg({-x^2 \Lambda^2 \over 4} \Bigg) + 2
\gamma_E \Bigg] \Bigg\}~,
\eea
where $\Lambda$ is the scale parameter and we choose it as the factorization
scale, i.e., $\Lambda=1~GeV$ (more about the discussion of this point can be
found in \cite{R9524}). 

The matrix elements $\lla \gamma(q) \vel \bar{q}_1
(x_1) \Gamma q(x_2) \ver o \rra$ are determined in terms of the photon Da's
in the following way \cite{R9525}:
\bea
\label{e9511}
&& \lla \gamma(q) \vert  \bar q(x) \sigma_{\mu \nu} q(0) \vert  0 \rra  =
-i e_q \qq (\varepsilon_\mu q_\nu - \varepsilon_\nu q_\mu) 
\int_0^1 du e^{i \bar u qx} 
\Bigg(\chi \varphi_\gamma(u) + {x^2 \over 16} \mathbb{A}  (u) \Bigg) \nnb \\  
&& - {i\over 2(qx)}  e_q \qq \Bigg[x_\nu \Bigg(\varepsilon_\mu - 
q_\mu {\varepsilon x \over qx}\Bigg) - 
x_\mu \Bigg(\varepsilon_\nu - q_\nu {\varepsilon x\over q x}\Bigg) \Bigg]
\int_0^1 du e^{i \bar u q x} h_\gamma(u) \nnb \\
&& \lla \gamma(q) \vert  \bar q(x) \gamma_\mu q(0) \vert 0 \rra  =
e_q f_{3 \gamma} \Bigg(\varepsilon_\mu - q_\mu {\varepsilon x\over q x} 
\Bigg) \int_0^1 du e^{i \bar u q x} \psi^v(u) \nnb \\
&& \lla \gamma(q) \vert \bar q(x) \gamma_\mu \gamma_5 q(0) \vert 0 \rra  =
- {1\over 4} e_q f_{3 \gamma} \epsilon_{\mu \nu \alpha \beta } 
\varepsilon^\nu q^\alpha x^\beta \int_0^1 du e^{i \bar u q x} 
\psi^a(u) \nnb \\
&& \lla \gamma(q) | \bar q(x) g_s G_{\mu \nu} (v x) q(0) \vert 0 \rra =
-i e_q \qq \Bigg(\varepsilon_\mu q_\nu - \varepsilon_\nu q_\mu \Bigg) 
\int {\cal D}\alpha_i e^{i (\alpha_{\bar q} + v \alpha_g) q x} 
{\cal S}(\alpha_i) \nnb \\
&& \lla \gamma(q) | \bar q(x) g_s \tilde G_{\mu \nu} i \gamma_5 (v x) 
q(0) \vert 0 \rra = -i e_q \qq \Bigg(\varepsilon_\mu q_\nu - 
\varepsilon_\nu q_\mu \Bigg)  \int {\cal D}\alpha_i e^{i (\alpha_{\bar q} + 
v \alpha_g) q x} \tilde {\cal S}(\alpha_i) \nnb \\
&& \lla \gamma(q) \vert \bar q(x) g_s \tilde G_{\mu \nu}(v x) 
\gamma_\alpha \gamma_5 q(0)  \vert 0 \rra =
e_q f_{3 \gamma} q_\alpha (\varepsilon_\mu q_\nu - \varepsilon_\nu q_\mu) 
\int {\cal D}\alpha_i e^{i (\alpha_{\bar q} + v \alpha_g) q x} {\cal A}
(\alpha_i) \nnb \\ 
&& \lla \gamma(q) \vert \bar q(x) g_s G_{\mu \nu}(v x) i \gamma_\alpha q(0) 
\vert 0 \rra = e_q f_{3 \gamma} q_\alpha (\varepsilon_\mu q_\nu - 
\varepsilon_\nu q_\mu) \int {\cal D}\alpha_i e^{i (\alpha_{\bar q} + 
v \alpha_g) q x} {\cal V}(\alpha_i) \nnb \\ 
&& \lla \gamma(q) \vert \bar q(x) \sigma_{\alpha \beta} g_s 
G_{\mu \nu}(v x) q(0) \vert 0 \rra  = e_q \qq \Bigg\{
\Bigg[\Bigg(\varepsilon_\mu - q_\mu {\varepsilon x\over q x}\Bigg)
\Bigg(g_{\alpha \nu} - {1\over qx} (q_\alpha x_\nu + q_\nu x_\alpha)
\Bigg) q_\beta \nnb \\  
&& - \Bigg(\varepsilon_\mu - q_\mu {\varepsilon x\over q x}\Bigg)
\Bigg(g_{\beta \nu} - {1\over qx} (q_\beta x_\nu + q_\nu x_\beta)
\Bigg) q_\alpha \nnb \\  
&& - \Bigg(\varepsilon_\nu - q_\nu {\varepsilon x\over q x}\Bigg)
\Bigg(g_{\alpha \mu} - {1\over qx} (q_\alpha x_\mu + q_\mu x_\alpha)
\Bigg) q_\beta \nnb \\ && + \Bigg(\varepsilon_\nu - q_\nu {\varepsilon 
x\over qx}\Bigg)\Bigg( g_{\beta \mu} - {1\over qx} (q_\beta x_\mu + 
q_\mu x_\beta)\Bigg) q_\alpha \Bigg] \int {\cal D}\alpha_i 
e^{i (\alpha_{\bar q} + v \alpha_g) qx} {\cal T}_1(\alpha_i) \nnb \\ 
&& + \Bigg[\Bigg(\varepsilon_\alpha - q_\alpha {\varepsilon x\over 
qx}\Bigg) \Bigg(g_{\mu \beta} - {1\over qx}(q_\mu x_\beta + q_\beta 
x_\mu)\Bigg)  q_\nu \nnb \\ 
&& - \Bigg(\varepsilon_\alpha - q_\alpha {\varepsilon x\over qx}
\Bigg) \Bigg(g_{\nu \beta} - {1\over qx}(q_\nu x_\beta + q_\beta 
x_\nu)\Bigg)  q_\mu \nnb \\ 
&& - \Bigg(\varepsilon_\beta - q_\beta {\varepsilon x\over qx}\Bigg)
\Bigg(g_{\mu \alpha} - {1\over qx}(q_\mu x_\alpha + q_\alpha 
x_\mu)\Bigg) q_\nu \nnb \\ 
&& + \Bigg(\varepsilon_\beta - q_\beta {\varepsilon x\over qx}\Bigg)
\Bigg(g_{\nu \alpha} - {1\over qx}(q_\nu x_\alpha + q_\alpha x_\nu) 
\Bigg) q_\mu \Bigg] \int {\cal D} \alpha_i e^{i (\alpha_{\bar q} + 
v \alpha_g) qx} {\cal T}_2(\alpha_i) \nnb \\ 
&& + {1\over qx} (q_\mu x_\nu - q_\nu x_\mu) (\varepsilon_\alpha 
q_\beta - \varepsilon_\beta q_\alpha) \int {\cal D} \alpha_i 
e^{i (\alpha_{\bar q} + v \alpha_g) qx} {\cal T}_3(\alpha_i) \nnb \\
&& + {1\over qx} (q_\alpha x_\beta - q_\beta x_\alpha)
(\varepsilon_\mu q_\nu - \varepsilon_\nu q_\mu)
\int {\cal D} \alpha_i e^{i (\alpha_{\bar q} + v \alpha_g) qx} 
{\cal T}_4(\alpha_i) \Bigg\}~,
\end{eqnarray}
where $\chi$ is the magnetic susceptibility of the quarks, $\varphi_\gamma(u)$ is the leading
twist 2, $\psi^v(u)$, $\psi^a(u)$, ${\cal A}$ and ${\cal V}$ are the twist 3 and $h_\gamma(u)$,
$\mathbb{A}$, ${\cal T}_i$ ($i=1,~2,~3,~4$) are the twist 4 photon distribution amplitudes.
The explicit expressions of these Da's are given in \cite{R9524}. The
integral measure ${\cal D} \alpha_i $ is defined as
\bea
\label{e9512}
{\cal D} \alpha_i = \int_0^1 d\alpha_g \int_0^1 d\alpha_q \int_0^1
d\alpha_{\bar{q}}~ \delta(1-\alpha_g - \alpha_q  - \alpha_{\bar{q}})~.
\eea

Substituting Eqs. (\ref{e9510}) and (\ref{e9511}) into Eq. (\ref{e9509}) and
performing integration over $x$, one can obtain the expression for the
correlation function $\Pi_{\mu\nu}^{th}$ in the momentum space.
Matching two different representations $\Pi^{ph}$ and $\Pi^{th}$ of the 
correlation function via the dispersion relation and applying double Borel
transformations on the variables $p^2$ and $(p+q)^2$, which suppresses
higher states and continuum contributions, we obtain the sum rules for the
form factors. Separating the coefficients of the structures $q_\mu
\varepsilon_\nu$ and $(\varepsilon p)q_\mu q_\nu$, which are free of contact
term contributions, we obtain the following sum rules for mesons containing
$u$ and $d$ quarks, i.e., for $\rho^+$ and $a_1^+$ mesons:
\bea
\label{e9513}
F_{\cal M}(0) \es {1\over f_i^2 m_i^2} e^{m_i^2/M^2} 
\Bigg\{
\mp {1\over 48 M^4} \GG \Big[e_u m_d \uu \mathbb{A}(\bar{u}_0) \nnb \\
\ek 2 e_u m_d \uu
\Big( u_0 \tilde{i}_3(h_\gamma,1) + \tilde{i}_3(h_\gamma,u-\bar{u}_0)
\Big) \nnb \\
\ek e_d m_u \dd \Big( \mathbb{A}(u_0) - 2 u_0 \tilde{i}_3^\prime(h_\gamma,1) +
2 \tilde{i}_3^\prime(h_\gamma,u-u_0) \Big) \Big]~, \nnb \\
\ar {1\over 24 M^2} \Big[ 2 m_0^2 \Big(\mp 3 e_u m_u \dd + 
2 e_u m_d u_0 \dd \pm 3 e_d m_d \uu - 2 e_d m_u u_0 \uu \Big) \nnb \\   
\kmp e_u m_d \GG \uu \chi \varphi_\gamma(\bar{u}_0) \pm e_d m_u \GG \dd \chi
\varphi_\gamma(u_0) \Big] \nnb \\
\ek {1\over 8 \pi^2} (e_d-e_u) M^4 (3 + 4 u_0) E_1(s/M^2) \nnb \\
\ar E_0(s/M^2) M^2 \chi \Big( \mp e_u m_d \uu
\varphi_\gamma(\bar{u}_0) \pm e_d m_u \dd \varphi_\gamma(u_0) \Big) \nnb \\
\ar {1\over 4} \Big[
E_0(s/M^2) M^2 f_{3\gamma} \Big(4 e_u i_2({\cal A},\bar{v}) -
4 e_u i_2({\cal V},\bar{v}) - 4 e_d i_2^\prime({\cal A},v)
- 4 e_d i_2^\prime({\cal V},v) \nnb \\
\ar 4 e_u \tilde{i}_3(\psi^v,1)
- 4 e_d \tilde{i}_3^\prime(\psi^v,1) - e_u \psi^a(\bar{u}_0)
+ e_d \psi^a(u_0) + 4 e_u u_0 \psi^v(\bar{u}_0) \nnb \\
\ek 4 e_d u_0 \psi^v(u_0)
+ e_u u_0 \psi^{a\prime}(\bar{u}_0) + e_d u_0 \psi^{a\prime}(u_0) \Big)
\Big] \nnb \\
\ar \uu \Bigg[ {1\over 2} e_d \Big(\pm 2 m_d + m_u\Big) + e_u m_d
\Big(\pm i_1({\cal S},1)     
\pm i_1(\tilde{{\cal S}},1) \pm i_1({\cal T}_1,1) \nnb \\
\kpm i_1({\cal T}_2,1) - i_1({\cal T}_3,1) - i_1({\cal T}_4,1)
+ u_0 \tilde{i}_3(h_\gamma,1) +    
\tilde{i}_3(h_\gamma,u-\bar{u}_0) \Big) \Bigg] \nnb \\
\ar \dd \Bigg[ - {1\over 2} e_u \Big(m_d \pm 2 m_u \Big)
\mp e_d m_u \Big(i_1^\prime({\cal S},1) + i_1^\prime(\tilde{{\cal S}},1)
- i_1^\prime({\cal T}_1,1) \nnb \\
\ek i_1^\prime({\cal T}_2,1) +
i_1^\prime({\cal T}_3,1) + i_1^\prime({\cal T}_4,1)
\mp u_0 \tilde{i}_3^\prime(h_\gamma,1) 
\pm \tilde{i}_3^\prime(h_\gamma,u-u_0) \Big) \Bigg]
\Bigg\}~, \nnb \\  \nnb \\
F_{\cal C}(0) + F_{\cal D}(0) \es {1\over f_i^2 m_i^2} e^{m_i^2/M^2} 
\Bigg\{{1\over 3 M^4} m_0^2 u_0^2 (e_u m_d \dd - e_d m_u \uu) \nnb \\
\ar {1\over M^2} \Big[
- e_u m_d u_0 \dd + e_d m_u u_0 \uu \nnb \\
\kpm 4 e_u m_d u_0 \uu \Big( i_0({\cal T}_1,1) + i_0({\cal T}_2,1)
- i_0({\cal T}_3,1) - i_0({\cal T}_4,1) \Big) \nnb \\
\kpm
4 e_d m_u u_0 \dd \Big( i_0^\prime({\cal T}_1,1) + i_0^\prime({\cal T}_2,1) - 
i_0^\prime({\cal T}_3,1) - i_0^\prime({\cal T}_4,1) \Big) \Big] \nnb \\
\ar {1 \over 4 \pi^2} M^2 u_0 (e_d - e_u) E_0(s/M^2) (3 - 4 u_0) \nnb \\
\ar 2 f_{3\gamma} \Big[ -e_u i_1({\cal A},1) + e_u i_1({\cal V},1-2 v) +
2 e_u u_0 \tilde{i}_2(\psi^v,1) \nnb \\
\ar e_d \Big(i_1^\prime ({\cal A},1) - i_1^\prime ({\cal V},1-2 v) -
2 u_0 \tilde{i}_3^\prime(\psi^v,1) \Big) \Big]
\Bigg\}~,
\eea
where the upper(lower) sign corresponds to the light--vector(axial--vector)
meson, $\chi$ is the magnetic susceptibility, and the continuum contributions 
are described by the function
\bea
\label{e9514}
E_n(x) = 1 - e^{-x} \sum_{i=0}^n {x^i \over i!}~,
\eea
where $x=s_0/M^2$ and $s_0$ is the continuum threshold. The Borel parameters
$M_1$ and $M_2$ are taken to be equal to each other, i.e., $M_1^2=M_2^2=2
M^2$, since we deal with a single meson, we have then
\bea
\label{e9515}
u_0={M_1^2 \over M_1^2+M_2^2}={1\over 2}~.
\eea

The functions $i_n$, $i_n^\prime$, $\tilde{i}_n$ and $\tilde{i}_n^\prime$ are
defined as
\bea
\label{nolabel}
i_0(\phi,f(v)) \es \int {\cal D}\alpha_i \int_0^1 dv
\phi(\alpha_{\bar{q}},\alpha_q,\alpha_g) f(v) \theta(k-u_0)~, \nnb \\
i_0^\prime(\phi,f(v)) \es \int {\cal D}\alpha_i \int_0^1 dv
\phi(\alpha_{\bar{q}},\alpha_q,\alpha_g) f(v) \theta(k^\prime-u_0)~, \nnb \\
i_1(\phi,f(v)) \es \int {\cal D}\alpha_i \int_0^1 dv
\phi(\alpha_{\bar{q}},\alpha_q,\alpha_g) f(v) \delta(k-u_0)~, \nnb \\
i_1^\prime(\phi,f(v)) \es \int {\cal D}\alpha_i \int_0^1 dv
\phi(\alpha_{\bar{q}},\alpha_q,\alpha_g) f(v) \delta(k^\prime-u_0)~, \nnb \\
i_2(\phi,f(v)) \es \int {\cal D}\alpha_i \int_0^1 dv
\phi(\alpha_{\bar{q}},\alpha_q,\alpha_g) f(v) \delta^\prime(k-u_0)~, \nnb \\
i_2^\prime(\phi,f(v)) \es \int {\cal D}\alpha_i \int_0^1 dv
\phi(\alpha_{\bar{q}},\alpha_q,\alpha_g) f(v) \delta^\prime(k^\prime-u_0)~, \nnb \\
\tilde{i}_3(\phi,f(u)) \es \int_0^{\bar{u}_0} du \phi(u) f(u)~, \nnb \\
\tilde{i}_3^\prime(\phi,f(u)) \es \int_{u_0}^1 du \phi(u) f(u)~, \nnb
\eea
where $k = \alpha_q + \alpha_g \bar{v}$ and $k^\prime = 
\alpha_{\bar{q}} + \alpha_g v$. The result for the $K^{\ast 0}$ 
($K^{\ast +}$) meson can be
obtained from Eq. (\ref{e9513}) by making the replacement $u\rar s$
($d \rar s$).

\section{Numerical analysis}

In this section we calculate the magnetic and quadrupole moments of the
light--vector and axial--vector mesons. The values of the input parameters
we use in our numerical analysis are,
$\uu(1~GeV) = \dd(1~GeV)= -(0.243)^3~GeV^3$, $\sp(1~GeV) = 0.8 \uu(1~GeV)$, 
$m_0^2(1~GeV) = 0.8$ \cite{R9523},
$\chi(1~GeV)=-4.4~GeV^{-2}$ \cite{R9526}, $\Lambda = 300~MeV$ and 
$f_{3 \gamma} = - 0.0039~GeV^2$ \cite{R9525},
$m_\rho=0.77~GeV$, $f_\rho=0.215~GeV$, 
$m_{K^\ast}=0.892~GeV$, $f_{K^\ast}=0.217~GeV$,
$m_{a_1}=1.260~GeV$, $f_{a_1}=0.200~GeV$.

The photon Da's entering the sum rules are \cite{R9525}:
\bea
\label{nolabel}
\varphi_\gamma(u) \es 6 u \bar u \left( 1 + \varphi_2(\mu) 
C_2^{\frac{3}{2}}(u - \bar u) \right)
\nnb \\
\psi^v(u) \es 3 \left(3 (2 u - 1)^2 -1 \right)+\frac{3}{64} 
\left(15 w^V_\gamma - 5 w^A_\gamma\right) 
      \left(3 - 30 (2 u - 1)^2 + 35 (2 u -1)^4 \right)
\nnb \\
\psi^a(u) \es \left(1- (2 u -1)^2\right)\left(5 (2 u -1)^2 
-1\right) \frac{5}{2} 
	\left(1 + \frac{9}{16} w^V_\gamma - 
\frac{3}{16} w^A_\gamma \right)
\nnb \\ 
{\cal A}(\alpha_i) \es 360 \alpha_q \alpha_{\bar q} \alpha_g^2 
		\left(1 + w^A_\gamma \frac{1}{2} (7 \alpha_g - 3)\right)
\nnb \\
{\cal V}(\alpha_i) \es 540 w^V_\gamma (\alpha_q - \alpha_{\bar q}) 
\alpha_q \alpha_{\bar q}
				\alpha_g^2
\nnb \\
h_\gamma(u) \es - 10 \left(1 + 2 \kappa^+\right) C_2^{\frac{1}{2}}
(u - \bar u) 
\nnb \\ 
\mathbb{A}(u) \es 40 u^2 \bar u^2 \left(3 \kappa - \kappa^+ +1\right) 
\nnb \\ && +
		8 (\zeta_2^+ - 3 \zeta_2) \left[u \bar u 
(2 + 13 u \bar u) \right. 
\nnb \\ && + \left.
                2 u^3 (10 -15 u + 6 u^2) \ln(u) + 2 \bar u^3 
(10 - 15 \bar u + 6 \bar u^2) 
		\ln(\bar u) \right]
\nnb \\
{\cal T}_1(\alpha_i) \es -120 (2 \zeta_2 + \zeta_2^+)
(\alpha_{\bar q} - \alpha_q) 
		\alpha_{\bar q} \alpha_q \alpha_g 
\nnb \\ 
{\cal T}_2(\alpha_i) \es 30 \alpha_g^2 (\alpha_{\bar q} - \alpha_q) 
	\left((\kappa - \kappa^+) + (\zeta_1 - \zeta_1^+)(1 - 2\alpha_g) + 
	\zeta_2 (3 - 4 \alpha_g)\right)
\nnb \\
{\cal T}_3(\alpha_i) \es - 120 (3 \zeta_2 - \zeta_2^+)
(\alpha_{\bar q} -\alpha_q) 
		\alpha_{\bar q} \alpha_q \alpha_g 
\nnb \\
{\cal T}_4(\alpha_i) \es 30 \alpha_g^2 (\alpha_{\bar q} - \alpha_q) 
	\left((\kappa + \kappa^+) + (\zeta_1 + \zeta_1^+)
(1 - 2\alpha_g) + 
	\zeta_2 (3 - 4 \alpha_g)\right)
\eea
The values of the constant parameters appearing in the Da's are
\cite{R9525}:
 $\varphi_2(1~GeV) = 0$, $w^V_\gamma = 3.8 \pm 1.8$,
$w^A_\gamma = -2.1 \pm 1.0$, $\kappa = 0.2$, $\kappa^+ = 0$, $\zeta_1 = 0.4$, $\zeta_2 = 0.3$,
$\zeta_1^+ = 0$ and $\zeta_2^+ = 0$.

The Borel mass is the artificial parameter of the sum rules and the
physically measurable quantities should be independent of them. For this
reason, we must find ``working region" of $M^2$, where the values of the
magnetic and quadrupole moments are, practically, independent of $M^2$.
In order to find the upper bound of $M^2$, we require that the contributions
continuum and higher states be less than $30\%$ of the total results. In
other words, the Borel parameter $M^2$ should not be too large in order to
guarantee that the above--mentioned contributions are exponentially
suppressed. Moreover, the lower bound of $M^2$ is determined through the
following argument. The Borel parameter could not be too small to satisfy
the validity of the OPE of the correlation function
near the light cone in the Euclidean region, since higher twist
contributions are proportional to $1/M^2$. As a result of these
constraints, the working regions of $M^2$ are determined to be:
\bea
\label{nolabel}
&& 0.8~GeV^2 \le M^2 \le 1.8~GeV^2~~(\rho~\mbox{\rm meson})~, \nnb \\
&& 1.0~GeV^2 \le M^2 \le 2.0~GeV^2~~(K^\ast~\mbox{\rm meson})~, \nnb \\
&& 1.5~GeV^2 \le M^2 \le 3.0~GeV^2~~(a_1~\mbox{\rm meson})~. \nnb
\eea
For the continuum threshold $s_0$, we choose, $s_0^{(\rho)}=1.7~GeV^2$,
$s_0^{(K^\ast)}=2.0~GeV^2$ and $s_0^{(a_1)}=3.0~GeV^2$.

The numerical results in analysis of the sum rules for the magnetic and
quadrupole moments, which is the main task of the present work, are
presented in Table (1).  

\newcommand{\rb}[1]{\raisebox{0.75ex}[0pt]{#1}}
\newcommand{\rbb}[1]{\raisebox{0.0ex}[0pt]{#1}}
\newcommand{\lb}[1]{\raisebox{-0.75ex}[0pt]{#1}}
\begin{table}
\renewcommand{\arraystretch}{1.4}
\addtolength{\arraycolsep}{1.5pt}
$$
\begin{array}{|c|c|c|c|c|c|c|}
\hline 
                                    &
\multicolumn{6}{|c|}{\mbox{\Large{$\mathbf{\mu}$}}~(\mbox{in}~e/2 m_i)} \\
\cline{2-7}\rbb{\bf Meson}
               & \lb{present}    & \lb{\cite{R9509}}   & \lb{\cite{R9510}}    & \lb{\cite{R9516}}    & \lb{covariant}  &  \lb{light cone} \\
               & \rb{work}       &                     &                 &                 &
\rb{quark model \cite{R9527}}    &  \rb{quark model \cite{R9512}} \\ \hline
\rho^+         &  2.4 \pm 0.4    &  2.01            &  2.2       &  2.7         & 2.14        &  1.92      \\ \hline
K^{\ast +}     &  2.0 \pm 0.4    &  2.23            &  2.08      &  2.36        & \mbox{---}  & \mbox{---} \\ \hline
K^{\ast 0}     &  0.28 \pm 0.04  &  -0.26           &  -0.08     &  -0.06       & \mbox{---}  & \mbox{---} \\ \hline
a_1^+          &  3.8 \pm 0.6    &   \mbox{---}     & \mbox{---} & 2            & \mbox{---}  & \mbox{---} \\ \hline
\end{array}
$$
\caption{The magnetic moments of light--vector and axial--vector mesons (in
units of $e/2 m_i$)}
\renewcommand{\arraystretch}{1}
\addtolength{\arraycolsep}{-1.5pt}
\end{table}
\begin{table}[h]
\renewcommand{\arraystretch}{1.4}
\addtolength{\arraycolsep}{1.5pt}
$$
\begin{array}{|c|c|c|c|c|c|c|}
\hline 
                                    &
\multicolumn{6}{|c|}{\mbox{\Large{$\mathbf{\cal D}$}}~(\mbox{in}~e/m_i^2)} \\
\cline{2-7}\rbb{\bf Meson}
               & \lb{present}    & \lb{\cite{R9509}}   & \lb{\cite{R9510}}    & \lb{\cite{R9516}}    & \lb{covariant}  &  \lb{light cone} \\
               & \rb{work}       &                     &                 &                 &
\rb{quark model \cite{R9527}}       &  \rb{quark model \cite{R9512}} \\ \hline
\rho^+         &  -0.85 \pm 0.15    & -0.41        &  \mbox{---} & \mbox{---} & -0.79      & -0.043     \\ \hline
K^{\ast +}     &  -0.8 \pm 0.15      & -0.38        &  \mbox{---} & \mbox{---} & \mbox{---} & \mbox{---} \\ \hline
K^{\ast 0}     &  -0.008 \pm 0.0004 &  0.01        &  \mbox{---} & \mbox{---} & \mbox{---} & \mbox{---} \\ \hline
a_1^+          &  -0.9 \pm 0.3      &   \mbox{---} &  \mbox{---} & \mbox{---} & \mbox{---} & \mbox{---} \\ \hline
\end{array}
$$
\caption{The quadrupole moments of light--vector and axial--vector mesons (in  
units of $e/m_i^2$)}
\renewcommand{\arraystretch}{1}
\addtolength{\arraycolsep}{-1.5pt}
\end{table}

The error in our predictions come from the variations in the Borel parameter,
$s_0$ and uncertainties from the nonperturbative input parameters in the
Da's of photon. 
 
For a comparison, in these Tables, we also present predictions of the
other approaches. We see from Tables (1) and (2) that within the limits of
errors, the predictions of different approaches on the magnetic moment of
the $\rho$ and $K^{\ast +}$ mesons are very close to each other.

As we have already noted, the magnetic moment of the $\rho$ meson is studied
within the framework of LCSR in \cite{R9515}, which is slightly different
from our prediction. This small difference can be attributed to the Da's
that are neglected in \cite{R9515}.

Our prediction of $\mu_{K^{\ast 0}}$ is more or less in agreement with the
prediction of \cite{R9505}, except its sign, while it drastically differs
from the predictions of \cite{R9509} and \cite{R9516}. Additionally, 
our result on the magnetic moment of $a_1^+$ differs considerably from 
the one given in \cite{R9516}. The situation on the quadrupole moments can
be summarized as follows. Our result on for ${\cal D}_{\rho^+}$ coincides
with the prediction of \cite{R9527}. But our results for ${\cal
D}_{K^{\ast +}}$ and ${\cal D}_{K^{\ast 0}}$ are almost twice as larger
compared to the predictions of the other approaches.

Our final remark is that we have also calculated the magnetic and
quadrupole moments of the $\rho^0$, $\omega$ and $\phi$ mesons, which are 
all equal to zero, as expected.  

In conclusion, we have calculated the magnetic and quadrupole moments of the
light-- and axial--vector mesons within the frame work of LCSR method. Our
results for the magnetic moments of the $\rho^+$ and $K^{\ast +}$ mesons are
in agreement with the predictions of the other approaches. Our prediction on
the magnetic moment of the $K^{\ast 0}$ meson, except its sign, confirms the
prediction of \cite{R9510}, while drastic differences are observed in
comparison to the other approaches. Note also that, our prediction of the
magnetic moment of axial--vector meson $a_1^+$, is 1.5--2 times larger
compared to that given in \cite{R9516}.

In regard to the quadrupole moments of the light--vector mesons, we conclude
that our predictions are approximately 2 times larger in comparison to the
other models, except the one predicted by \cite{R9523}, which is in close
agreement with ours.

\end{document}